\documentclass[reprint,showpacs,superscriptaddress,preprintnumbers,amsmath,amssymb,aps,prd,nofootinbib]{revtex4-1}
\usepackage[utf8]{inputenc}
\usepackage{bm}
\usepackage{hyperref}
\usepackage{graphicx}

\newcommand{\total}{\mathop{}\!\mathrm{d}}
\newcommand{\eqend}[1]{\,\mathrm{#1}}
\newcommand{\mathi}{\mathrm{i}}
\newcommand{\mathe}{\mathrm{e}}
\renewcommand{\vec}[1]{\bm{#1}}

\newcommand{\abs}[1]{{\left\lvert{#1}\right\rvert}}

\newcommand{\bigo}[1]{\mathcal{O}\left({#1}\right)}

\newenvironment{equations}[1][]{\subequations\ifx\relax#1\relax\else\label{#1}\fi\align\ignorespaces}{\endalign\endsubequations}

\allowdisplaybreaks
\usepackage{xspace}
\newcommand{\etal}{et\,al.\xspace}
\newcommand{\ie}{i.\,e.\xspace}
\newcommand{\eg}{e.\,g.\xspace}

\begin{document}

\title{Mapping AdS to dS spaces and back}

\author{Adriana Di Dato}
\email{adidato@ffn.ub.es}
\affiliation{Departament de F{\'\i}sica Fonamental, Institut de Ci{\`e}ncies del Cosmos (ICC), Universitat de Barcelona (UB), C/ Mart{\'\i} i Franqu{\`e}s 1, 08028 Barcelona, Spain}

\author{Markus B. Fr{\"o}b}
\email{mfroeb@itp.uni-leipzig.de}
\affiliation{Departament de F{\'\i}sica Fonamental, Institut de Ci{\`e}ncies del Cosmos (ICC), Universitat de Barcelona (UB), C/ Mart{\'\i} i Franqu{\`e}s 1, 08028 Barcelona, Spain}
\affiliation{Institut f{\"u}r Theoretische Physik, Universit{\"a}t Leipzig, Br{\"u}derstra{\ss}e 16, 04103 Leipzig, Germany}

\date{\today}
\begin{abstract}
We derive a map between Einstein spaces of positive and negative curvature, including scalar matter. Starting from a space of positive curvature with some dimensions compactified on a sphere and analytically continuing the number of compact dimensions, we obtain a space of negative curvature with a compact hyperbolic subspace, and vice versa. Prime examples of such spaces are de Sitter (dS) and anti-de Sitter (AdS) space, as well as black hole spacetimes with (A)dS asymptotics and perturbed versions thereof, which play an important role in holography. This map extends work done by Caldarelli \etal, who map asymptotically AdS spaces to Ricci-flat ones. A remarkable result is that the boundary of asymptotically AdS spaces is mapped to a brane in the bulk of de Sitter, and perturbations near the AdS boundary are sourced by a stress tensor confined to this brane. We also calculate the Brown-York stress tensor for the perturbed AdS metric, which turns out to be the negative of the stress tensor on the de Sitter brane. The map can also be used as a solution generator, and we obtain a Kerr/AdS solution with hyperbolic horizon from a known Kerr/dS one.
\end{abstract}


\maketitle

\section{Introduction}
\label{intro}

Since the discovery of the AdS/CFT correspondence, a concrete realization of the holographic idea that theories with gravity can be described by theories without gravity in one dimension less, a lot of effort has been invested in the study of this and other holographic dualities. An area in which a holographic duality would be very useful is for the description of the early Universe, especially for inflation. However, the geometry of the Universe at that time is close to de Sitter (dS) space~\cite{mukhanov}, and also today the measured cosmological constant is positive~\cite{planck}, so that the AdS/CFT correspondence is not directly applicable. A dS/CFT correspondence has been proposed by Strominger~\cite{strominger2001,strominger2001b} (see also Refs.~\cite{nojiri2001,ghezelbash2002,klemm2002,spradlin2002,anninos2011,maldacena2011}), but the boundary CFT can be nonunitary and contain complex conformal weights (\eg, for sufficiently massive scalars in dS). Another approach to use holography in inflation has been put forward by McFadden and Skenderis~\cite{mcfadden2009,mcfadden2010,mcfadden2011,bzowski2012}, where correlators are calculated using the standard AdS/CFT correspondence and then analytically continued to complex momenta to obtain results for de Sitter spacetime. This construction has been tested to give the right predictions for correlators of gravitons and inflaton perturbations, which are both massless fields, but it is not assured that it works for massive fields as well.

Recently, a map between solutions of the Einstein equations with negative cosmological constant and Ricci-flat solutions was derived by Caldarelli \etal~\cite{caldarelli2013a,caldarelli2013b} using generalized dimensional reduction, a diagonal Kaluza-Klein (KK) dimensional reduction~\cite{duff1986,pope} followed by an analytic continuation in the number of dimensions~\cite{gouteraux2012,caldarelli2013a,caldarelli2013b,smolic2013}, which especially includes a map between asymptotically AdS and asymptotically flat spacetimes. This map does not involve an analytic continuation in the complex plane, but instead rests on a suitable compactification of some coordinates in each space. Furthermore, it can be used to understand how to set up holography for Minkowski spacetime, and in this context a remarkable fact was discovered: the holographic stress tensor in AdS is mapped to a brane situated at $r=0$ in Minkowski spacetime, which serves as the source for the metric perturbations. This is in contrast to previous works that, in analogy with the AdS case, studied holography at various boundaries of flat space~\cite{deboer2003,arcioni2003,mann2005}.

In this paper, we generalize the construction of Refs.~\cite{caldarelli2013a,caldarelli2013b} to solutions of the Einstein equations with positive and negative cosmological constants, including matter in the form of a scalar field --- a map that can bring AdS to dS and vice versa. The possibility of such a map had already been mentioned in Ref.~\cite{caldarelli2013b}, but only the matter-free reduced action in the Jordan frame was calculated there. Our paper is organized as follows: First, we derive the map using a diagonal KK dimensional reduction of the action. Reducing also the higher-dimensional Einstein equations (which leads to the same result), we then show that the reduction ansatz is consistent. Afterwards, we give some examples: the maps between empty AdS and dS spaces, between black holes with AdS/dS asymptotics and for perturbations near the boundary of AdS, which are relevant for holography. For this last case, we calculate the Brown-York stress tensor and compare with holographic expectations. By mapping a known Kerr/dS black hole, we also find a (most probably new) solution for a rotating black hole in AdS with a hyperbolic horizon, showing the feasibility of using the map as a solution generator.

For the metric and curvature tensors, we use the ``+++'' convention of Ref.~\cite{mtw}. Capital latin indices denote coordinates in the higher-dimensional space before reduction, and lowercase latin (greek) indices denote coordinates in the reduced (compact) directions. Quantities which refer to (asymptotic) dS space are indicated by a prime, while quantities without a prime either are general or refer to (asymptotic) AdS space.

\section{Deriving the map}
\label{map}

In this section, we show how the map can be derived by a diagonal KK dimensional reduction of a higher-dimensional system, once directly at the level of the action and once by reducing the higher-dimensional Einstein equations. We start from a $(n+\nu)$-dimensional spacetime that is a solution of the Einstein equations with a cosmological constant (which can be either positive or negative) and matter. Of these coordinates, $\nu$ will be compactified, with the size of the compactification determined by a scalar field $\phi$ (a dilaton) that only depends on the $n$ reduced coordinates. That is, we start from a metric
\begin{equation}
\total \bar{s}^2 = \bar{g}_{MN} \total X^M \total X^N \eqend{,} \qquad M,N = 0, \ldots, n+\nu-1
\end{equation}
that solves the Einstein equations with matter
\begin{equation}
\bar{G}_{MN} + \Lambda \bar{g}_{MN} = 8 \pi G^{n+\nu}_\text{N} \bar{T}_{MN}
\end{equation}
in $n+\nu$ dimensions. We perform dimensional reduction by taking the ansatz
\begin{equation}
\label{map_productmetric}
\total \bar{s}^2 = \mathe^{2\alpha\phi(x)} \total s^2 + \mathe^{2\beta\phi(x)} \total \sigma^2
\end{equation}
with the $n$-dimensional reduced metric
\begin{equation}
\total s^2 = g_{ab}(x) \total x^a \total x^b
\end{equation}
and the $\nu$-dimensional compact metric
\begin{equation}
\total \sigma^2 = \gamma_{\alpha\beta}(y) \total y^\alpha \total y^\beta \eqend{.}
\end{equation}
We take the metric of the compact space $\gamma_{\alpha\beta}$ to be fixed, while the reduced metric $g_{ab}$ is dynamical and, like the scalar $\phi$, only depends on the coordinates $x^a$. The parameters $\alpha$ and $\beta$ are constants and can be chosen at will; however, $\beta$ cannot be zero for a consistent reduction, as can be seen later on from the reduced equations~\eqref{eom}.

\subsection{Generalized dimensional reduction of the action}
\label{dimreduxaction}

The $(n+\nu)$-dimensional action is the Einstein-Hilbert action with cosmological constant $\Lambda$ and a free, canonically normalized scalar field $\chi$
\begin{equation}
\label{dimreduxaction_action}
S = \int \left[ \frac{\bar{R} - 2 \Lambda}{16\pi G^{n+\nu}_\text{N}} - \frac{1}{2} \bar{g}^{AB} \partial_A \bar{\chi} \partial_B \bar{\chi} \right] \sqrt{-\bar{g}} \total^{n+\nu} X \eqend{,}
\end{equation}
where $G^{n+\nu}_\text{N}$ is Newton's constant in $n+\nu$ dimensions. For the dimensional reduction, the scalar $\bar{\chi}$ is taken to only depend on the reduced coordinates, and to simplify the formulas we rescale it
\begin{equation}
\chi = \sqrt {16\pi G^{n+\nu}_\text{N}} \bar{\chi} \eqend{.}
\end{equation}
Calculating the curvature tensors for the ansatz~\eqref{map_productmetric}, which is done in appendix~\ref{app_curvature}, and substituting them into the action~\eqref{dimreduxaction_action}, we obtain (after integration by parts and ignoring surface terms)
\begin{equation}
\begin{split}
S &= \frac{1}{16\pi G^{n+\nu}_\text{N}} \int \mathe^{\left[ (n-2) \alpha + \nu \beta \right] \phi} \bigg[ R - 2 \mathe^{2 \alpha \phi} \Lambda \\
&\quad+ \mathe^{2 (\alpha-\beta) \phi} R[\gamma] - \frac{1}{2} \nabla^a \chi \nabla_a \chi \\
&\quad+ \Big( (n-1) (n-2) \alpha^2 + 2 (n-1) \nu \alpha \beta \\
&\qquad\quad+ \nu (\nu-1) \beta^2 \Big) \nabla^a \phi \nabla_a \phi \bigg] \sqrt{-g} \sqrt{\gamma} \total^n x \total^\nu y \eqend{.}
\end{split}
\end{equation}
In this expression, $R[\gamma]$ is the Ricci scalar of the compact metric $\gamma_{\alpha\beta}$, and $\nabla$ is the covariant derivative with respect to the reduced metric $g_{ab}$. Setting $\alpha = 0$, $\beta = 1/\nu$ and $\chi = 0$, we recover the matter-free reduced action in the Jordan frame derived in Ref.~\cite{caldarelli2013b}.

Now take the compact space to be an Einstein space which has $R_{\mu\nu}[\gamma] = k (\nu-1) H^2 \gamma_{\mu\nu}$, where $H$ is a constant with dimensions of inverse length related to the radius of the compact space (\eg, for a sphere, the radius would be $H^{-1}$). The constant $k$ takes the values $\pm 1$, and the compact space has volume $V^k_\nu \propto H^{-\nu}$. We express the cosmological constant as $\Lambda = \lambda (n+\nu-1)(n+\nu-2)/(2\ell^2)$ with $\lambda = \pm 1$, and a constant $\ell$ with dimensions of length (\eg, in pure AdS, $\ell$ is the AdS radius). Integrating out the compact coordinates, we get
\begin{equation}
\label{dimreduxaction_reduxaction}
\begin{split}
S &= \frac{V^k_\nu}{16\pi G^{n+\nu}_\text{N}} \int \bigg[ R - \lambda \frac{(n+\nu-1)(n+\nu-2)}{\ell^2} \mathe^{2 \alpha \phi} \\
&\quad+ \nu (\nu-1) k H^2 \mathe^{2 (\alpha-\beta) \phi} - \frac{1}{2} \nabla^a \chi \nabla_a \chi \\
&\quad+ \Big( (n-1) (n-2) \alpha^2 + 2 (n-1) \nu \alpha \beta \\
&\qquad\quad+ \nu (\nu-1) \beta^2 \Big) \nabla^a \phi \nabla_a \phi \bigg] \mathe^{\left[ (n-2) \alpha + \nu \beta \right] \phi} \sqrt{-g} \total^n x \eqend{.}
\end{split}
\end{equation}

To construct a map between a space which is a solution for $\Lambda > 0$ and one for $\Lambda < 0$, we perform this reduction twice, with different internal spaces. On one hand, we consider the reduced action $S$ with $\Lambda < 0$ (and thus $\lambda = -1$), and on the other hand a second reduced action $S'$ (denoted by primes) with $\Lambda' > 0$ (and thus $\lambda' = +1$). The actions $S$ and $S'$ are proportional to each other,
\begin{equation}
\label{action_prop}
S = \frac{G^{n'+\nu'}_\text{N}}{V^{k'}_{\nu'}} \frac{V^k_\nu}{G^{n+\nu}_\text{N}} S' \eqend{,}
\end{equation}
if and only if $k=-1$, $k'=+1$ and
\begin{equations}[dimreduxaction_map]
\ell &= 1/H' \eqend{,} & H &= 1/\ell' \eqend{,} \\
\alpha &= \alpha' - \beta' \eqend{,} & \beta &= - \beta' \eqend{,} \\
n &= n' \eqend{,} & \nu &= 2-n'-\nu' \eqend{.}
\end{equations}
This is consistent with the AdS/Ricci-flat correspondence~\cite{caldarelli2013a,caldarelli2013b}, where, however, the constants $\alpha$ and $\beta$ were fixed (choosing a specific frame and a canonical normalization for the scalar field).

That is, given a solution of the Einstein equations with negative cosmological constant of the form
\begin{equation}
\label{dimreduxaction_mapads}
\total \bar{s}^2 = \mathe^{2\alpha\phi(x)} g^{(n)}_{ab} \total x^a \total x^b + \mathe^{2\beta\phi(x)} \gamma^{(\nu)-}_{\alpha\beta} \total y^\alpha \total y^\beta \eqend{,}
\end{equation}
where we have shown explicitly the dimensions of the metrics and denoted the negative curvature of the compact space by a minus sign, and a scalar field $\chi$, the line element
\begin{equation}
\label{dimreduxaction_mapds}
\begin{split}
\total \tilde{s}^2 &= \mathe^{2\alpha'\phi(x)} g^{(n')}_{ab} \total x^a \total x^b + \mathe^{2\beta'\phi(x)} \gamma^{(\nu')+}_{\alpha\beta} \total y^\alpha \total y^\beta \\
&= \mathe^{2(\alpha-\beta)\phi(x)} g^{(n)}_{ab} \total x^a \total x^b \\
&\qquad+ \mathe^{-2\beta\phi(x)} \gamma^{(2-n-\nu)+}_{\alpha\beta} \total y^\alpha \total y^\beta
\end{split}
\end{equation}
with the \emph{same} metric $g^{(n)}_{ab}$ and scalar field $\phi$ as well as the \emph{same} scalar $\chi$ gives a solution of the Einstein equations with positive cosmological constant. This map is valid in general dimensions, and while the dimensions of the reduced spaces are the same, the dimension of the compact space changes. For sufficiently large $n$ and $\nu$, $\nu'$ will be negative and must be analytically continued to a positive value. This poses no problem as long as no factors of $1/\nu'$ (or similar) appear.

It is important to note that any explicit factors of $n$, $\nu$, $\ell$ or $H$ appearing in the metric and the scalar fields (or $\alpha$ and $\beta$) must be identified as well using~\eqref{dimreduxaction_map}, so that one needs to know the solution for arbitrary $\nu$ (since $n' = n$, one may fix $n$). If one wants to work in a specific frame (Einstein or Jordan), one may fix $\alpha$ or $\beta$, but this is not necessary for the map. Furthermore, exchanging primed and unprimed quantities in~\eqref{dimreduxaction_map}, we see that the map works likewise both ways.

Since the actions are equal up to an overall constant~\eqref{action_prop}, the equations of motion for the reduced space and the scalar field are the same. However, we need to check the consistency of the reduction, \ie, that the Einstein equations that follow from the reduced action~\eqref{dimreduxaction_reduxaction} can be obtained by reducing the equations that follow from the starting action~\eqref{dimreduxaction_action}. This will be done in the next section.

\subsection{Reduction of the Einstein equations}
\label{dimreduxeqns}

The Einstein equations that follow from the original $(n+\nu)$-dimensional action~\eqref{dimreduxaction_action} are
\begin{equation}
\label{eom_general}
\begin{split}
&\bar{R}_{AB} - \frac{1}{2} \bar{R} \bar{g}_{AB} + \Lambda \bar{g}_{AB} \\
&\quad= \frac{1}{2} \left[ \bar{\nabla}_A \chi \bar{\nabla}_B \chi - \frac{1}{2} \bar{g}_{AB} \bar{\nabla}^C \chi \bar{\nabla}_C \chi \right] \eqend{,}
\end{split}
\end{equation}
and the scalar field equation reads $\bar{\nabla}^A \bar{\nabla}_A \chi = 0$.
Using the product space metric ansatz~\eqref{map_productmetric}, they can be decomposed using the formulas from appendix~\ref{app_curvature}. Imposing (as in the last section) that the compact metric is Einstein with Ricci tensor $R_{\mu\nu}[\gamma] = k (\nu-1) H^2 \gamma_{\mu\nu}$ and taking $\Lambda = \lambda (n+\nu-1)(n+\nu-2)/(2\ell^2)$, we obtain after some algebraic manipulations
\begin{equations}[eom]
\begin{split}
&R_{mn} - \lambda (n+\nu-1)/\ell^2 \mathe^{2 \alpha \phi} g_{mn} - \alpha g_{mn} \nabla^a \nabla_a \phi \\
&\quad- \left[ (n-2) \alpha + \nu \beta \right] \nabla_m \nabla_n \phi \\
&\quad+ \left[ (n-2) \alpha^2 + 2 \nu \alpha \beta - \nu \beta^2 \right] \nabla_m \phi \nabla_n \phi \\
&\quad- \alpha \left[ (n-2) \alpha + \nu \beta \right] g_{mn} \nabla^a \phi \nabla_a \phi = \frac{1}{2} \nabla_m \chi \nabla_n \chi \eqend{,}
\end{split} \\
\begin{split}
&\beta \Big[ \nabla^a \nabla_a \phi + \left[ (n-2) \alpha + \nu \beta \right] \nabla^a \phi \nabla_a \phi \Big] \\
&\quad+ \left[ \lambda (n+\nu-1)/\ell^2 - k (\nu-1) H^2 \mathe^{- 2 \beta \phi} \right] \mathe^{2 \alpha \phi} = 0 \eqend{,}
\end{split} \\
&\nabla^a \nabla_a \chi = \left[ (n-2) \alpha - \nu \beta \right] \left( \nabla^a \phi \right) \nabla_a \chi \eqend{.}
\end{equations}
These are exactly the equations that follow by varying the reduced action~\eqref{dimreduxaction_reduxaction}. We thus conclude that the reduction is consistent. Here we also see why the restriction $\beta \neq 0$ is important: for $\beta = 0$, the second equation does not give any restriction on the dilaton, but instead relates the sizes of the extended and the compact space.

In the Einstein frame, we have $(n-2) \alpha + \nu \beta = 0$, and the equations reduce to the simpler ones
\begin{equations}
\begin{split}
&R_{mn} - \lambda (n+\nu-1)/\ell^2 \mathe^{2 \alpha \phi} g_{mn} - \alpha g_{mn} \nabla^a \nabla_a \phi \\
&\quad+ \nu \beta (\alpha-\beta) \nabla_m \phi \nabla_n \phi = \frac{1}{2} \nabla_m \chi \nabla_n \chi \eqend{,}
\end{split} \\
\begin{split}
&\beta \nabla^a \nabla_a \phi = \\
&\quad- \left[ \lambda (n+\nu-1)/\ell^2 - k (\nu-1) H^2 \mathe^{- 2 \beta \phi} \right] \mathe^{2 \alpha \phi} \eqend{,}
\end{split}\\
&\nabla^a \nabla_a \chi = 0 \eqend{.}
\end{equations}
Note that while $\alpha$, $\beta$ and $\nu$ individually change under the map~\eqref{dimreduxaction_map}, ``being in the Einstein frame'' is a condition that is preserved, as can be easily seen from the term-by-term comparison of the corresponding actions. However, it is almost always easier to work with $\alpha = 0$ or $\pm 1$ and $\beta = \pm 1$, as we will do in the following.

\section{Applications}
\label{applications}

In this section, we apply the map derived above to concrete examples. First, we give a short introduction to compact hyperbolic spaces, which are an example of Riemannian Einstein manifolds of constant negative curvature, and which are the simplest space of negative Ricci curvature to use in the map. Afterwards, we show that we can map pure dS to pure AdS. The next subsection then treats small perturbations on top of AdS, which arise in holography in the AdS/CFT correspondence, in order to explore a possible dS/CFT correspondence via our map. Lastly, we show how to map the Schwarzschild-dS black hole to the Schwarzschild-AdS black hole, and map a rotating Kerr/dS black hole to AdS, obtaining a (probably new) Kerr/AdS black hole solution with hyperbolic horizon.

\subsection{Compact hyperbolic spaces}
\label{chs}

Hyperbolic spaces are the analogue of AdS in Riemannian geometry, in the same way that the sphere is the Riemannian analogue of dS. It is well known that the $\nu$-dimensional unit sphere can be defined by embedding it into a $(\nu+1)$-dimensional flat Euclidean space known as ambient space, where it arises as the submanifold
\begin{equation}
\label{chs_sphere_embedding}
\delta_{AB} X^A X^B = 1 \eqend{,} \qquad A,B = 1, \ldots, \nu+1 \eqend{.}
\end{equation}
The metric of the sphere is then the induced metric obtained by restricting the flat ambient metric $\delta_{AB}$ to this submanifold. In the same way, hyperbolic spaces (of unit radius) are obtained from an ambient space with flat Lorentzian metric $\eta_{AB}$ as the submanifold
\begin{equation}
\label{chs_embedding}
\eta_{AB} X^A X^B = - 1 \eqend{.}
\end{equation}
Choosing
\begin{equation}
X^1 = \frac{4 \delta_{\alpha\beta} y^\alpha y^\beta - 1}{4 y^1} \eqend{,} \qquad X^A = \frac{y^\alpha}{y^1} \eqend{,} \ A = \alpha = 2, \ldots, \nu
\end{equation}
and solving equation~\eqref{chs_embedding} for $X^0$, one obtains the induced metric
\begin{equation}
\gamma_{\alpha\beta} = \eta_{AB} \frac{\total X^A}{\total y^\alpha} \frac{\total X^B}{\total y^\beta} = \frac{\delta_{\alpha\beta}}{(y^1)^2} \eqend{.}
\end{equation}
In these coordinates, it is clear that hyperbolic space is the Riemannian analogue of AdS (identifying $y^1$ with $r$, where $r$ is the radial coordinate). Another coordinate system which will be more suited for the purposes of the map later on is obtained by choosing
\begin{equation}
X^1 = \sinh y_1 \cos y_2
\end{equation}
and taking the coordinates $X^A$ for $A = 2, \ldots, \nu$ to be spherical coordinates with radius $\sinh y_1 \sin y_2$. This choice gives the induced metric
\begin{equation}
\label{chs_globalhyper}
\gamma_{\alpha\beta} \total y^\alpha \total y^\beta = \total y_1^2 + \sinh^2 y_1 \total \Omega_{\nu-1}^2 \eqend{.}
\end{equation}
Spaces which do not have unit radius are then obtained by simply multiplying the metric by the (constant) radius.

One now has to compactify this space, which is done by taking the quotient by a discrete subgroup of isometries (which are the isometries of the ambient space that leave invariant the submanifold~\eqref{chs_embedding}). Of course, the local metric does not change under this compactification, and one easily calculates that
\begin{equation}
R_{abcd}[\gamma] = - H^2 ( \gamma_{ac} \gamma_{bd} - \gamma_{ad} \gamma_{bc} )
\end{equation}
for all compact hyperbolic spaces (CHSs). An example of such a compactification can easily be given: take the two-dimensional hyperbolic space with metric
\begin{equation}
\total s^2 = \frac{\total x^2 + \total y^2}{(Hy)^2} \eqend{.}
\end{equation}
The isometry group of this metric is formed by the transformations
\begin{equation}
(x+\mathi y) \to \frac{a (x + \mathi y) + b}{c (x + \mathi y) + d}
\end{equation}
with $ad-bc = 1$ (the Möbius transformations), as one can easily check. A discrete subgroup of this group is the modular group, where the parameters $a$, $b$, $c$ and $d$ are restricted to be integers. One then identifies points which are mapped one into the other by the action of this subgroup, which \eg includes $x \to x + k$, $k \in \mathbb{Z}$ (for $a = d = 1$, $b = k$ and $c = 0$). A fundamental domain for this group action is given by points which have $x^2+y^2 \geq 1$ and $\abs{x} \leq \frac{1}{2}$, and the CHS is obtained by identifying the borders, just like the torus can be obtained by identifying the sides of a rectangle. The volume of this space is given by
\begin{equation}
V^-_2 = \int_{-\frac{1}{2}}^\frac{1}{2} \int_{\sqrt{1-x^2}}^\infty (Hy)^{-2} \total y \total x = \frac{\pi}{3 H^2} \eqend{,}
\end{equation}
which is finite, showing that this CHS really is compact. In higher dimensions, there are plenty of CHSs~\cite{thurston1982,kaloper2000,nasri2002}, and so there is no problem using them in our map.

\subsection{AdS/dS spacetimes}
\label{adsds}

Of course, the simplest examples for the map are empty dS/AdS spaces. Taking $\chi = 0$, it turns out to be easier to start from the dS side, where the $(n'+\nu')$-dimensional metric (in the Poincar\'e patch) takes the form
\begin{equation}
\label{adsds_dsmetric}
\total \tilde{s}^2 = \frac{(\ell')^2}{\eta^2} \left( - \total \eta^2 + \total \vec{x}^2_{n'-2} + \total r^2 + r^2 \total \Omega_{\nu'}^2 \right) \eqend{,}
\end{equation}
where we compactified $\nu'$ coordinates on a sphere, with metric $\total \Omega_{\nu'}^2$. Comparing with the general formula~\eqref{dimreduxaction_mapds}, the most economic choice is to take $\alpha' = 0$ and $\beta' = -1$, which means $\alpha = \beta = 1$. The reduced metric then reads (using the identification~\eqref{dimreduxaction_map} for the second equality)
\begin{equation}
\label{bg_ads_metric}
\begin{split}
g^{(n)}_{ab} \total x^a \total x^b &= \frac{(\ell')^2}{\eta^2} \left( - \total \eta^2 + \total \vec{x}^2_{n'-2} + \total r^2 \right) \\
&= \frac{1}{(H\eta)^2} \left( - \total \eta^2 + \total \vec{x}^2_{n-2} + \total r^2 \right) \eqend{,}
\end{split}
\end{equation}
and the scalar field is given by
\begin{equation}
\label{bg_ads_phi}
\phi = \ln \left( \frac{\eta}{H' \ell' r} \right) = \ln \left( \frac{H \ell \eta}{r} \right)
\end{equation}
(recall that the compact space was taken to be of radius $1/H'$ in the map~\eqref{dimreduxaction_map}, which needs to be compensated by the scalar field since there is no $H'$ in the metric~\eqref{adsds_dsmetric}). The map tells us that the metric obtained from equation~\eqref{dimreduxaction_mapads}
\begin{equation}
\label{bg_ads_sol}
\begin{split}
\total \bar{s}^2 &= \mathe^{2\phi} g^{(n)}_{ab} \total x^a \total x^b + \mathe^{2\phi} \total \sigma^2_\nu \\
&= \frac{\ell^2}{r^2} \left( - \total \eta^2 + \total \vec{x}^2_{n-2} + \total r^2 + \eta^2 \total \Upsilon^2_\nu \right) \eqend{,}
\end{split}
\end{equation}
with $\total \Upsilon^2_\nu$ the $\nu$-dimensional line element of a CHS of unit radius, is a solution of the Einstein equations with negative cosmological constant. To recover the metric of the Poincar\'e patch of AdS, we take the metric of the CHS in the form~\eqref{chs_globalhyper}
\begin{equation}
\label{upsilon}
\total \Upsilon^2_\nu = \total y_1^2 + \sinh^2 y_1 \total \Omega^2_{\nu-1}
\end{equation}
and perform the coordinate transformation
\begin{equation}
\label{coord_trafo}
\eta^2 = t^2 - \vec{z}_\nu^2 \eqend{,} \qquad \sinh^2 y_1 = \frac{\vec{z}_\nu^2}{t^2 - \vec{z}_\nu^2} \eqend{.}
\end{equation}
Then we obtain (analogous to the Milne universe~\cite{mukhanov})
\begin{equation}
- \total \eta^2 + \eta^2 \total \Upsilon^2_\nu = - \total t^2 + \total \abs{\vec{z}_\nu}^2 + \vec{z}_\nu^2 \total \Omega^2_{\nu-1} = - \total t^2 + \total \vec{z}_\nu^2 \eqend{,}
\end{equation}
so that the metric~\eqref{bg_ads_sol} reduces to
\begin{equation}
\label{ads_empty}
\total \bar{s}^2 = \frac{\ell^2}{r^2} \left( - \total t^2 + \total \vec{x}^2_{n-2} + \total r^2 + \total \vec{z}_\nu^2 \right)
\end{equation}
which is AdS in $n+\nu$ dimensions, with $r$ the radial/holographic coordinate. By compactifying $\nu'$ coordinates of de~Sitter space on a sphere and applying the map, we thus find AdS space.

An interesting feature of the map concerns the position of the AdS boundary, which is located at $r=0$. Since the extended metric does not change under the mapping, this surface is located in the bulk of dS, and has itself the geometry of a dS space: an $(n-1)$-dimensional dS brane. This happens in a similar manner in the AdS/Ricci-flat correspondence~\cite{caldarelli2013a, caldarelli2013b}, where the AdS boundary is mapped to a flat brane in the bulk of Minkowski spacetime, and we will discuss implications of this fact in Sec.~\ref{pert}, where we treat perturbations in AdS/dS.

Just like AdS, empty dS enjoys a conformal symmetry. Of special importance are dilatations and special conformal transformations. We undo the compactification in~\eqref{adsds_dsmetric}, writing
\begin{equation}
\label{uncompact}
\total r^2 + r^2 \total \Omega_{\nu'}^2 = \total \vec{r}_{\nu'+1}^2 \eqend{,}
\end{equation}
so that the metric takes the form
\begin{equation}
\label{dsmetric_decompact}
\total \tilde{s}^2 = \frac{(\ell')^2}{\eta^2} \left( - \total \eta^2 + \total \vec{x}^2_{n'-2} + \total \vec{r}_{\nu'+1}^2 \right) \eqend{.}
\end{equation}
The dilatations are given by
\begin{equation}
\eta \to \lambda \eta \eqend{,} \quad \vec{x} \to \lambda \vec{x} \eqend{,} \quad \vec{r} \to \lambda \vec{r} \eqend{.}
\end{equation}
The invariance of the metric under this transformation is clear. Defining $\vec{z} = \{\vec{x},\vec{r}\}$, special conformal transformations read
\begin{equations}
\eta \to \eta - 2 \eta (\vec{b}\vec{z}) \eqend{,} \\
\vec{z} \to \vec{z} + \vec{b} (\vec{z}^2) - 2 \vec{z} (\vec{b}\vec{z}) - \eta^2 \vec{b} \eqend{,}
\end{equations}
with $\vec{b}$ an infinitesimal constant vector. One can easily verify that~\eqref{dsmetric_decompact} is invariant. After the compactification, we are only interested in the transformation of the reduced part. We decompose therefore $\vec{b} = \{\vec{b}^x,\vec{b}^r\}$ and define $c \equiv (\vec{b}^r \vec{r})/r$. The transformation of the reduced coordinates then only depends on $\vec{b}^x$ and $c$, and we calculate
\begin{equations}
\eta &\to \eta - 2 \eta (\vec{b}^x\vec{x} + c r) \eqend{,} \\
\vec{x} &\to \vec{x} + \vec{b}^x (\vec{x}^2 + r^2 - \eta^2) - 2 \vec{x} (\vec{b}^x\vec{x} + c r) \eqend{,} \\
r &\to r + c (\vec{x}^2 + r^2 - \eta^2) - 2 r (\vec{b}^x\vec{x} + c r) \eqend{.}
\end{equations}
The reduced metric~\eqref{bg_ads_metric} is invariant under this transformation, but the dilaton~\eqref{bg_ads_phi} changes as
\begin{equation}
\phi \to \phi - \frac{c}{r} (\vec{x}^2 + r^2 - \eta^2) \eqend{.}
\end{equation}
We see that the compactification breaks the original conformal symmetry, but the resulting transformations can be seen as a generalized conformal structure --- the reduced metric is conformally invariant, but the dilaton introduces a scale in the theory. These transformations are, however, solution generating transformations, as can be checked from the equations~\eqref{eom}. Since the map brings solutions to solutions, the same transformations are valid in AdS space.

\subsection{Asymptotic AdS with perturbations}
\label{pert}

In the AdS/CFT correspondence, the large $N$ and large 't\,Hooft coupling limit of the conformal field theory corresponds to a weakly coupled gravity theory that can be described by supergravity in an asymptotically AdS space. The dictionary, the precise relation between these theories including renormalization, is known~\cite{balasubramanian1999,deharo2001}, and in this section we calculate how perturbations near the AdS boundary, which are relevant in this holographic dictionary, are mapped to perturbations around dS. Again, we treat vacuum solutions and take $\chi = 0$.

We therefore approach the mapping from the other direction: take $\alpha = 0$ and $\beta = 1$, and a reduced metric and dilaton of the form
\begin{equations}
g^{(n)}_{ab} &= \frac{\ell^2}{r^2} \left( \eta_{ab} + h_{ab}(\eta,\vec{x},r) \right) \\
\phi &= \ln \left( \frac{H \ell \eta}{r} \right) + \psi(\eta,\vec{x},r) \eqend{,}
\end{equations}
where we take the Fefferman-Graham gauge~\cite{fefferman}: $h_{ab}$ does not have components in the radial direction, and both $h_{ab}$ and $\psi$ vanish as $r \to 0$. Both $h_{ab}$ and $\psi$ can be considered as perturbations on top of the background AdS metric, and we retain the correct asymptotic behavior as $r \to 0$. The full metric then reads
\begin{equation}
\label{aads_pert_metric}
\total \bar{s}^2 = \frac{\ell^2}{r^2} \left[ \left( \eta_{ab} + h_{ab} \right) \total x^a \total x^b + \mathe^{2\psi} \eta^2 \total \Upsilon_\nu^2 \right] \eqend{,}
\end{equation}
with $\total \Upsilon_\nu^2$ the line element of a CHS of unit radius~\eqref{upsilon}. We thus leave the compact space unperturbed, and only vary its radius. For $h_{ab} = \psi = 0$, the map gives the same de~Sitter metric~\eqref{adsds_dsmetric}, showing that $\alpha$ and $\beta$ can be chosen freely and in a suitable way for the problem at hand.

Since our boundary metric is flat (we just have the Poincar\'e patch of AdS in the unusual coordinates~\eqref{coord_trafo}), the relevant corrections are of the form~\cite{deharo2001,caldarelli2013b} ($n+\nu \geq 4$)
\begin{equations}[holo_pert]
h_{ab} &= r^d h^{(d)}_{ab}(\eta,\vec{x}) + r^{d+2} h^{(d+2)}_{ab}(\eta,\vec{x}) + \bigo{r^{d+3}} \eqend{,} \\
\psi &= r^d \psi^{(d)}(\eta,\vec{x}) + r^{d+2} \psi^{(d+2)}(\eta,\vec{x}) + \bigo{r^{d+3}} \eqend{,}
\end{equations}
where we defined $d \equiv n+\nu-1$. The (reduced) Einstein equations~\eqref{eom} then give
\begin{equations}[eom_pert]
h^{(d)} &= - 2 \nu \psi^{(d)} \eqend{,} \\
h^{(d+2)} &= - 2 \nu \psi^{(d+2)} \eqend{,} \\
\eta \partial^m h^{(d)}_{mn} &= \nu h^{(d)}_{n0} - \delta^0_n h^{(d)} \eqend{,} \\
\begin{split}
2 (d+2) \eta^2 h^{(d+2)}_{mn} &= - \eta^2 \partial^2 h^{(d)}_{mn} + \nu \eta \partial_\eta h^{(d)}_{mn} \\
&\quad- 2 \nu \delta^0_{(m} h^{(d)}_{n)0} + 2 \delta^0_m \delta^0_n h^{(d)} \eqend{.}
\end{split}
\end{equations}
For $\nu = 0$ (\ie, no compact dimensions), we should recover perturbations around pure AdS, and we indeed obtain
\begin{equations}
h^{(d)} &= 0 \eqend{,} \\
\partial^m h^{(d)}_{mn} &= 0 \eqend{,} \\
2 (d+2) h^{(d+2)}_{mn} &= - \partial^2 h^{(d)}_{mn} \eqend{,}
\end{equations}
which are the well-known conditions for asymptotically AdS spaces with a flat boundary~\cite{deharo2001,caldarelli2013b}.

\begin{figure}[ht]
\centering
\includegraphics[width=\columnwidth]{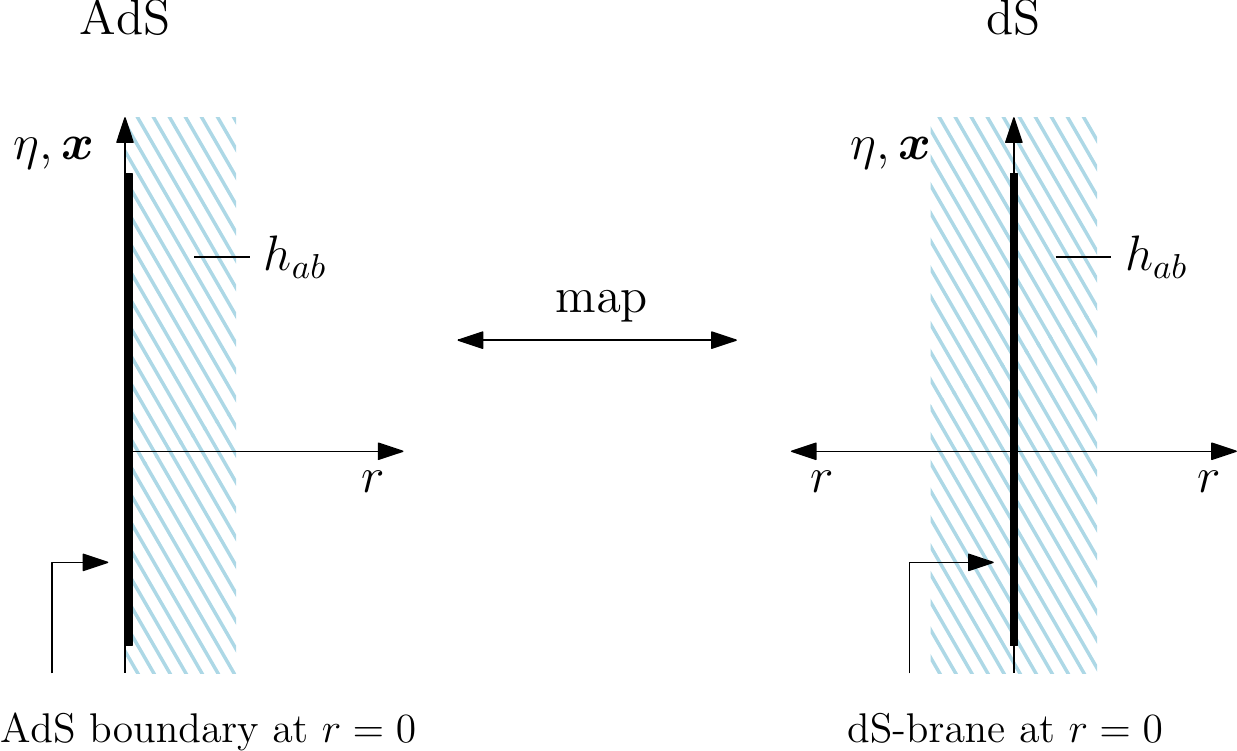}
\caption{The map for perturbations around (anti-)de~Sitter spacetime. On the AdS side, the perturbations live near the boundary at $r=0$, while the map puts them near a brane located in the bulk of dS at $r=0$.}
\label{fig-ads-ds-map}
\end{figure}

After performing the map~\eqref{dimreduxaction_mapds}, \eqref{dimreduxaction_map}, we obtain the de Sitter metric with perturbations of the form
\begin{equation}
\total \tilde{s}^2 = \frac{(\ell')^2}{\eta^2} \mathe^{-2\psi} \left[ \left( \eta_{ab} + h_{ab} \right) \total x^a \total x^b + r^2 \total \Omega_{\nu'}^2 \right] \eqend{.}
\end{equation}
However, due to the now singular factor $r^d = r^{1-\nu'}$ in the perturbations, they no longer fulfill the source-free Einstein equations~\eqref{eom}. On the de Sitter side after the map, we have $\alpha' = \beta' = -1$, so that the reduced Einstein equations (with a general source~\eqref{eom_stresstensor}) read
\begin{equations}
\begin{split}
&R_{mn} + g_{mn} \nabla^a \nabla_a \phi - (n'+\nu'-1)/(\ell')^2 \mathe^{-2 \phi} g_{mn} \\
&\quad+ (n'+\nu'-2) \left( \nabla_m \nabla_n \phi + \nabla_m \phi \nabla_n \phi - g_{mn} \nabla^a \phi \nabla_a \phi \right) \\
&\quad= 8 \pi G^{n'+\nu'}_\text{N} \left( T_{mn} - \frac{1}{(n'+\nu'-2)} g_{mn} T \right) \eqend{,}
\end{split} \\
\begin{split}
&- \nabla^a \nabla_a \phi + (n'+\nu'-2) \nabla^a \phi \nabla_a \phi - (\nu'-1) (H')^2 \\
&\quad+ (n'+\nu'-1)/(\ell')^2 \mathe^{-2 \phi} = \frac{8 \pi}{(n'+\nu'-2)} G^{n'+\nu'}_\text{N} T \eqend{,}
\end{split}
\end{equations}
where $T = g^{mn} T_{mn} = (H')^2 r^2 \eta^{mn} T_{mn}$. Putting the perturbations~\eqref{holo_pert} into these equations and using the conditions~\eqref{eom_pert} (taking care to replace $n \to n'$, $\nu \to 2 - n' - \nu'$ and $d \to 1 - \nu'$ according to the map), we obtain
\begin{equation}
\label{tmn_dsbrane}
\begin{split}
T_{mn} &= - \frac{(1-\nu')}{16 \pi G^{n'+\nu'}_\text{N}} h_{mn}^{(d)} r^{-\nu'} \delta(r) \\
&= - \left[ \left( \frac{1}{H'} \right)^{d-1} \!\! \frac{d}{16 \pi G^{d+1}_\text{N}} h_{mn}^{(d)} \right] \delta^{1+\nu'}(\vec{r}) \eqend{,}
\end{split}
\end{equation}
where we have ``uncompactified'' the compact coordinates as in~\eqref{uncompact}, and defined the $(d+1)$-dimensional Newton's constant as
\begin{equation}
G^{d+1}_\text{N} \equiv \frac{G^{n'+\nu'}_\text{N}}{V^+_{\nu'}} = \frac{G^{n'+\nu'}_\text{N}}{(H')^{-\nu'} \Omega_{\nu'}} \eqend{.}
\end{equation}
We thus see that the perturbations after the map are sourced by a stress tensor situated on a brane (with intrinsic de~Sitter geometry) located in the bulk of de~Sitter at $r=0$. This map is shown in Fig.~\ref{fig-ads-ds-map}.
Furthermore, if the conformal field theory living at the $(n+\nu-1)$-dimensional boundary of AdS before the compactification can be consistently reduced to an $n$-dimensional theory plus an additional scalar operator, the expectation value of the $n$-dimensional holographic stress tensor would be given by the term in brackets in~\eqref{tmn_dsbrane},
\begin{equation}
\label{tmn_cft}
\left\langle T^\text{CFT}_{mn} \right\rangle = \frac{d \ell^{d-1}}{16 \pi G^{d+1}_\text{N}} h_{mn}^{(d)}
\end{equation}
(taking into account the map: $n = n'$ and $H' = 1/\ell$). This is the same conclusion that has been reached in the AdS/Ricci-flat correspondence~\cite{caldarelli2013a,caldarelli2013b}: the (negative) dual stress tensor of the holographic CFT serves as a source for the metric perturbations after the map, with support on a brane situated in the bulk of Minkowski spacetime.

To reinforce these indications, we calculate the (subtracted and rescaled) quasilocal Brown-York stress tensor~\cite{brownyork,myers1999,balasubramanian1999,deharo2001} associated with a surface $r=\text{const.}$ in the metric~\eqref{aads_pert_metric}. The normal vector to this surface is given by
\begin{equation}
n^A = \frac{r}{\ell} \delta^A_r
\end{equation}
and normalized to $n^A n_A = 1$. The extrinsic curvature tensor $K_{AB}$ of the surface is defined by
\begin{equation}
K_{AB} = ( \delta_A^M - n_A n^M ) ( \delta_B^N - n_B n^N ) \nabla_M n_N \eqend{,}
\end{equation}
and we calculate
\begin{equation}
\nabla_M n_N = \frac{\ell}{r^2} \delta^r_M \delta^r_N + \frac{r}{2 \ell} \partial_r g_{MN}
\end{equation}
and from this
\begin{equations}
\begin{split}
K_{ab} &= - \frac{\ell}{r^2} \left( \eta_{ab} - \delta^r_a \delta^r_b \right) + \frac{\ell r}{2} \partial_r \left( \frac{1}{r^2} h_{ab} \right) \\
&= - \frac{\ell}{r^2} \left( \eta_{ab} - \delta^r_a \delta^r_b - \frac{(d-2)}{2} r^d h^{(d)}_{ab} \right) + \bigo{r^d} \eqend{,}
\end{split} \\
K_{a\beta} &= 0 \eqend{,} \\
\begin{split}
K_{\alpha\beta} &= \frac{\ell r}{2} \eta^2 \gamma^{(-1)}_{\alpha\beta} \partial_r \left( \frac{1}{r^2} \mathe^{2\psi} \right) \\
&= - \frac{\ell}{r^2} \eta^2 \gamma^{(-1)}_{\alpha\beta} \left( 1 - (d-2) r^d \psi^{(d)} \right) + \bigo{r^d} \eqend{,}
\end{split}
\end{equations}
where $\gamma^{(-1)}_{\alpha\beta}$ is the metric of a CHS of unit radius, $\total \Upsilon_\nu^2 = \gamma^{(-1)}_{\alpha\beta} \total y^\alpha \total y^\beta$. The trace of the extrinsic curvature follows as (using the conditions~\eqref{eom_pert})
\begin{equation}
K = g^{MN} K_{MN} = - \frac{d}{\ell} + \bigo{r^{d+2}} \eqend{.}
\end{equation}
The unsubtracted Brown-York stress tensor can be shown to be equal to~\cite{myers1999,balasubramanian1999}
\begin{equation}
\label{brownyork}
8 \pi G^{n+\nu}_\text{N} T^\text{BY}_{MN} = K_{MN} - K g_{MN} \eqend{,}
\end{equation}
and the counterterms that one needs to subtract from the stress tensor to make it well defined as $r \to 0$ depend on the dimension. For dimensions up to four, they are given by~\cite{balasubramanian1999,deharo2001}
\begin{equation}
\label{brownyork_ct}
\begin{split}
8 \pi G^{n+\nu}_\text{N} T^\text{CT}_{MN} &= - \frac{(d-1)}{\ell} g_{MN} \\
&\quad- \frac{\ell}{d-2} \left( \mathcal{R}_{MN} - \frac{1}{2} \mathcal{R} g_{MN} \right) \eqend{,}
\end{split}
\end{equation}
where $\mathcal{R}_{MN}$ is the Ricci tensor of the induced boundary metric $g_{MN}(r=\text{const})$. In even dimensions, there is an additional term (related to the conformal anomaly) which we do not consider here. The Ricci tensor for the induced boundary metric can be found in appendix~\ref{app_boundary}, and using the perturbation~\eqref{holo_pert} and the conditions~\eqref{eom_pert}, we find up to corrections of order $\bigo{r^d}$
\begin{equations}
8 \pi G^{n+\nu}_\text{N} T^\text{BY}_{ab} &= \frac{\ell}{r^2} \left[ (d-1) \eta_{ab} + \frac{(3d-2)}{2} r^d h^{(d)}_{ab} \right] \eqend{,} \\
8 \pi G^{n+\nu}_\text{N} T^\text{BY}_{a\beta} &= 0 \eqend{,} \\
8 \pi G^{n+\nu}_\text{N} T^\text{BY}_{\alpha\beta} &= \frac{\ell}{r^2} \eta^2 \gamma^{(-1)}_{\alpha\beta} \left[ (d-1) + (3d-2) r^d \psi^{(d)} \right] \eqend{,} \\
8 \pi G^{n+\nu}_\text{N} T^\text{CT}_{ab} &= - (d-1) \frac{\ell}{r^2} \left( \eta_{ab} + r^d h^{(d)}_{ab} \right) \eqend{,} \\
8 \pi G^{n+\nu}_\text{N} T^\text{CT}_{a\beta} &= 0 \eqend{,} \\
8 \pi G^{n+\nu}_\text{N} T^\text{CT}_{\alpha\beta} &= - (d-1) \frac{\ell}{r^2} \eta^2 \left( 1 + 2 r^d \psi^{(d)} \right) \gamma^{(-1)}_{\alpha\beta} \eqend{.}
\end{equations}
Note that to leading order, only the first counterterm in~\eqref{brownyork_ct} contributes, while the others are of order $\bigo{r^d}$. The subtracted and rescaled stress tensor is then given by~\cite{deharo2001}
\begin{equation}
T^\text{SR}_{MN} = \lim_{r \to 0} \left[ \left( \frac{\ell}{r} \right)^{d-2} \left( T^\text{BY}_{MN} + T^\text{CT}_{MN} \right) \right] \eqend{,}
\end{equation}
and we finally obtain
\begin{equations}
T^\text{SR}_{ab} &= \frac{d \ell^{d-1} h^{(d)}_{ab}}{16 \pi G^{n+\nu}_\text{N}} \eqend{,} \\
T^\text{SR}_{a\beta} &= 0 \eqend{,} \\
T^\text{SR}_{\alpha\beta} &= 2 \eta^2 \gamma^{(-1)}_{\alpha\beta} \frac{d \ell^{d-1} \psi^{(d)}}{16 \pi G^{n+\nu}_\text{N}} \eqend{.}
\end{equations}
This is consistent with our earlier remarks around equation~\eqref{tmn_cft}, and shows explicitly what form the stress tensor expectation value in the dual CFT would have to take.

\subsection{Black holes}
\label{bh}

In the case of black objects, solutions for general $n$ are much more difficult to obtain. Since the AdS/Ricci-flat map has been used to study hydrodynamics of black branes, we would also like to apply the AdS/dS map to a (planar, vacuum) black brane, with the metric
\begin{equation}
\total \bar{s}^2 = \frac{\ell^2}{r^2} \left( - f(r) \total t^2 + \total \vec{x}^2_{n-2} + \frac{\total r^2}{f(r)} + \total \vec{z}_\nu^2 \right)
\end{equation}
and $f(r) = 1 - (r/b)^{n+\nu-1}$ with a constant $b$. However, if we try to compactify $\vec{z}_\nu$ using the coordinate transformation~\eqref{coord_trafo}, the dilaton depends on the compactified coordinates, in contrast to our initial ansatz.

If we start from the dS side, a (vacuum) black brane solution is not known. Nevertheless, we can consider the Schwarzschild-dS black hole with fixed $n' = 2$, since the dimension of the reduced space does not change under the map. This solution is given in static coordinates by~\cite{kottler,bazanski1986}
\begin{equation}
\total \tilde{s}^2 = - f(r) \total t^2 + \frac{\total r^2}{f(r)} + r^2 \total \Omega_{\nu'}^2 \eqend{,}
\end{equation}
with
\begin{equation}
f(r) = 1 - \left( \frac{r_\text{S}}{r} \right)^{\nu'-1} - \frac{r^2}{(\ell')^2} = 1 - \left( \frac{r}{r_\text{S}} \right)^{\nu+1} - H^2 r^2 \eqend{,}
\end{equation}
and where $r_\text{S}$ is the Schwarzschild radius of the black hole, while $\ell'$ is the dS radius. Again, we already used the identification~\eqref{dimreduxaction_map} for the second equality. Comparison with~\eqref{dimreduxaction_mapds}, taking $\alpha' = 0$ and $\beta' = -1$ and thus $\alpha = \beta = 1$, gives us
\begin{equations}
g^{(n)}_{ab} \total x^a \total x^b &= - f(r) \total t^2 + \frac{\total r^2}{f(r)} \eqend{,} \\
\phi &= - \ln (H'r) = - \ln(r/\ell) \eqend{,}
\end{equations}
and the corresponding AdS metric obtained via the map~\eqref{dimreduxaction_mapads} reads
\begin{equation}
\total \bar{s}^2 = \frac{\ell^2}{r^2} \left( - f(r) \total t^2 + \frac{\total r^2}{f(r)} + H^{-2} \total \Upsilon_\nu^2 \right) \eqend{.}
\end{equation}
Changing coordinates to $z = \ell/(H r)$ and $t = \tau/(H\ell)$, we obtain
\begin{equation}
\total \bar{s}^2 = - \bar{f}(z) \total \tau^2 + \frac{\total z^2}{\bar{f}(z)} + z^2 \total \Upsilon_\nu^2
\end{equation}
with
\begin{equation}
\label{sads}
\bar{f}(z) = \frac{z^2}{\ell^2} f(\ell/(Hz)) = - 1 - \left( H r_\text{S} \right)^{-(\nu+1)} \left( \frac{\ell}{z} \right)^{\nu-1} + \frac{z^2}{\ell^2} \eqend{,}
\end{equation}
which is the metric for an AdS black hole with hyperbolic horizon geometry~\cite{birmingham1998,emparan1999}.

What happens to the horizons? We concentrate on the case of a small black hole, where $r_\text{S} \ll \ell'$. The black hole horizon is situated at $r_\text{BH} \approx r_\text{S}$, while the cosmological horizon is at $r_\text{CH} \approx \ell'$. The map gives $\ell' = 1/H$, so that we have $H r_\text{S} \ll 1$. After the coordinate transformation, the black hole horizon is mapped to $z_\text{BH} \approx \ell/(H r_\text{S}) \gg \ell$, and the cosmological horizon to $z_\text{CH} \approx \ell$. Plugging these values into $\tilde{f}(z)$, we see that for the black hole horizon we have $\tilde{f}(z_\text{BH}) \approx 0$ (since we can neglect the $-1$ in comparison with the huge term $( H r_\text{S} )^{-2}$), but for the cosmological horizon we obtain $\tilde{f}(z_\text{CH}) \approx - ( H r_\text{S} )^{-(\nu+1)} \neq 0$. This can be understood from the map: the term $- ( r_\text{S}/\ell' )^{\nu'-1}$, which was negligible at the cosmological horizon for positive $\nu'$, became $- ( H r_\text{S} )^{-(\nu+1)}$, which is large for positive $\nu$.

We see that the black hole horizons are mapped to each other, while the cosmological horizon disappears because of the analytic continuation in the number of dimensions.

Another class of black hole solutions which are interesting to analyze are rotating ones. Kerr/dS black holes have been constructed in higher dimensions with any number of rotation parameters~\cite{gibbonslupagepope2004}, but to show examples of the map one rotation parameter is enough. We use the metric given in Ref.~\cite{dehghani2002}, which describes a rotating black hole with mass parameter $M$ and (one) angular momentum parameter $a$. This metric reads
\begin{equation}
\begin{split}
\total \tilde{s}^2 &= - \frac{\Delta_r}{\rho^2} \left( \total t - \frac{a}{\Xi} \sin^2\theta \total \phi \right)^2 + \frac{\rho^2}{\Delta_r} \total r^2 + \frac{\rho^2}{\Delta_\theta} \total\theta^2 \\
&\quad+ \frac{\Delta_\theta \sin^2\theta}{\rho^2} \bigg( a \total t - \frac{r^2+a^2}{\Xi} \total \phi \bigg)^2 + r^2 \cos^2\theta \total \Omega_{\nu'}^2 \eqend{,}
\end{split}
\end{equation}
where
\begin{equations}
\Delta_r &= (r^2+a^2) \left( 1 - \frac{r^2}{(\ell')^2} \right) - 2 M r^{3-\nu'} \eqend{,} \\
\Delta_\theta &= 1 + \frac{a^2}{(\ell')^2} \cos^2\theta \eqend{,} \\
\Xi &= 1 + \frac{a^2}{(\ell')^2} \eqend{,} \\
\rho^2 &= r^2 + a^2 \cos^2\theta \eqend{.}
\end{equations}
The reduction proceeds in the same way as before, and again taking $\alpha' = 0$ and $\beta' = -1$, we have
\begin{equations}[kerrmapped]
g_{ab} \total x^a \total x^b &= \total \tilde{s}^2 - r^2 \cos^2\theta \total \Omega_{\nu'}^2 \eqend{,} \\
\phi &= - \ln ( H' r \cos\theta ) = - \ln\left( \frac{r}{\ell} \cos\theta \right) \eqend{.}
\end{equations}
The mapped rotating black hole in AdS is then given by
\begin{equation}
\begin{split}\label{KerrAdS}
\total \bar{s}^2 &= \frac{\ell^2}{r^2 \cos^2\theta} \Big[ - \frac{\tilde \Delta_r}{\rho^2} \left( \total t - \frac{a}{\tilde\Xi} \sin^2 \theta \total\phi \right)^2 + \frac{\rho^2}{\tilde\Delta_r} \total r^2 \\
&\quad+ \frac{\rho^2}{\tilde \Delta_\theta} \total\theta^2 + \frac{\tilde\Delta_\theta \sin^2\theta}{\rho^2} \left( a \total t - \frac{r^2 + a^2}{\tilde\Xi} \total \phi \right)^2 \\
&\quad+ H^{-2} \total \Upsilon_\nu^2 \Big] \eqend{,}
\end{split}
\end{equation}
where
\begin{equations}
\tilde{\Delta}_r &= \left( r^2 + a^2 \right) \left( 1 - H^2 r^2 \right) - 2 M r^{3+\nu} \eqend{,} \\
\tilde{\Delta}_\theta &= 1 + H^2 a^2 \cos^2\theta \eqend{,} \\
\tilde{\Xi} &= 1 + H^2 a^2 \eqend{.}
\end{equations}
One can see that this metric is singular near $\theta = \pi/2$, which is due to the fact that the compactified space (the $\nu'$-sphere) has vanishing radius at that point, and thus gives a singular dilaton. Such singular dilatons have also been found in some cases of T-duality~\cite{dijkgraafverlinde1992,kiritsis1993}. Calculating, \eg, the Kretschmann scalar one finds, however, the completely regular result
\begin{equation}
R_{ABCD} R^{ABCD} = \frac{2 (\nu+3) (\nu+4)}{l^4} + \bigo{\theta - \frac{\pi}{2}} \eqend{.}
\end{equation}
Since also for $a \to 0$ the metric does not reduce to the Schwarzschild-AdS black hole~\eqref{sads}, it is thus possible that a suitable coordinate transformation exists which yields a manifestly regular metric also for $\theta = \pi/2$. We leave a detailed investigation for further study.

\section{Conclusions}

In this article, we have presented a map between Einstein spaces of negative and positive curvature, including a scalar field. In order to obtain such a map via generalized dimensional reduction, these spaces need to have the form of a direct product between an extended spacetime (the bulk) and a compact subspace, whose curvature has the same sign as the total space. Especially, spacetimes which are asymptotically AdS, with the subspace being a compact hyperbolic space, are mapped to a spacetime which is asymptotically de Sitter (deep in the bulk), with the transverse subspace a sphere. This map is a generalization of the AdS/Ricci-flat correspondence~\cite{caldarelli2013a,caldarelli2013b}, and we expect it to generalize to the case of additional matter fields such as gauge fields. Furthermore, nondiagonal reductions are probably possible, as well as the study of moduli of the internal space (note, however, that compact hyperbolic spaces do not possess massless shape moduli by the Mostow rigidity theorem~\cite{gromov1981,kaloper2000}). In general, the mapping is between solutions with different compact dimensions, and the number of compact dimensions must be analytically continued after the map to a positive value. One must therefore know the solution for a general compact dimension $\nu$, and it must be regular as $\nu \to 0$ for the continuation to be unambiguous. However, this does not seem to be a strong restriction in practice, as exemplified by the application of the map to empty AdS/dS, black hole spacetimes and perturbations on top of AdS/dS.

A very direct application of the map is as a solution generator, mapping known solutions of positive and negative curvature to each other in nontrivial ways as exemplified by the asymptotically dS/AdS rotating black holes, where the AdS solution is most probably new. Other contexts of study suggest themselves: for example, the AdS/Ricci-flat map has been used to study hydrodynamics of black branes and the Gregory-Laflamme instability~\cite{caldarelli2013a,caldarelli2013b,didato2013,gath2014}. Using the AdS/dS map derived in this paper, these considerations could be extended also to de Sitter spacetime.

An important fact (which applies in the same way in the AdS/Ricci-flat correspondence) concerns the mapping of the AdS boundary, which is sent to a brane in the bulk of dS. This brane has itself an intrinsic de~Sitter geometry, and supports a stress tensor which serves as the source of perturbations, and which is the negative of the Brown-York stress tensor in the perturbed AdS geometry. These perturbations are obtained by mapping perturbations near the boundary of AdS that encode holographic information from the AdS/CFT correspondence, and the stress tensor on the brane is compatible with what one would expect if the dual CFT at the AdS boundary can be consistently reduced over a compact hyperbolic space (for the AdS/Ricci-flat correspondence, the reduction over a torus is consistent~\cite{kanitscheider2008,kanitscheider2009} and the corresponding statement can be made). This discovery suggests that a putative holographic dual to de Sitter space is not to be found at infinity in analogy with the AdS case, but instead on such a brane.

\begin{acknowledgments}
We would like to thank Kostas Skenderis for his inspiring Christmas talk at the University of Barcelona which brought us to the study of this fascinating subject, and Roberto Emparan for very valuable discussions, suggestions and comments on a draft of this paper.

M. F. acknowledges financial support through a FPU scholarship No.~AP2010-5453. A. D. acknowledges financial support through a FPI scholarship No.~BES-2011-045401. Both authors acknowledge partial financial support by the Research Projects MEC FPA2007-66665-C02-02, FPA2010-20807-C02-02, CPAN CSD2007-00042, within the program Con\-so\-li\-der-Ingenio 2010, and AGAUR 2009-SGR-00168. M. F. also acknowledges financial support through ERC starting grant QC\&C 259562.
\end{acknowledgments}

\appendix

\section{Curvature tensors for a product metric}
\label{app_curvature}

For the product space metric~\eqref{map_productmetric}, one easily calculates the Christoffel symbols directly from the definition
\begin{equations}
\bar{\Gamma}^a_{bc} &= \Gamma^a_{bc} + \alpha \left( \delta^a_c \nabla_b + \delta^a_b \nabla_c - g_{bc} \nabla^a \right) \phi \eqend{,} \\
\bar{\Gamma}^a_{b\gamma} &= 0 \eqend{,} \\
\bar{\Gamma}^a_{\beta\gamma} &= - \beta \mathe^{2 (\beta-\alpha) \phi} \gamma_{\beta\gamma} \nabla^a \phi \eqend{,} \\
\bar{\Gamma}^\alpha_{bc} &= 0 \eqend{,} \\
\bar{\Gamma}^\alpha_{b\gamma} &= \beta \delta^\alpha_\gamma \partial_b \phi \eqend{,} \\
\bar{\Gamma}^\alpha_{\beta\gamma} &= \Gamma^\alpha_{\beta\gamma}[\gamma] \eqend{.}
\end{equations}
The curvature tensors follow as
\begin{equations}
\begin{split}
\bar{R}_{mn} &= R_{mn} - \left[ (n-2) \alpha + \nu \beta \right] \nabla_m \nabla_n \phi \\
&\quad+ \left[ (n-2) \alpha^2 + 2 \nu \alpha \beta - \nu \beta^2 \right] \nabla_m \phi \nabla_n \phi \\
&\quad- \alpha g_{mn} \Big[ \nabla^a \nabla_a \phi + \left[ (n-2) \alpha + \nu \beta \right] \nabla^a \phi \nabla_a \phi \Big] \eqend{,}
\end{split} \\
\bar{R}_{m\nu} &= \bar{R}_{\mu n} = 0 \eqend{,} \\
\begin{split}
\bar{R}_{\mu\nu} &= R_{\mu\nu} - \beta \mathe^{2 (\beta-\alpha) \phi} \gamma_{\mu\nu} \nabla^a \nabla_a \phi \\
&\quad- \beta \mathe^{2 (\beta-\alpha) \phi} \gamma_{\mu\nu} \left[ (n-2) \alpha + \nu \beta \right] \nabla^a \phi \nabla_a \phi \eqend{,}
\end{split} \\
\begin{split}
\mathe^{2 \alpha \phi} \bar{R} &= R + \mathe^{2 (\alpha-\beta) \phi} R[\gamma] - 2 \left[ (n-1) \alpha + \nu \beta \right] \nabla^a \nabla_a \phi \\
&\quad- \Big[ (n-1) (n-2) \alpha^2 + 2 (n-2) \nu \alpha \beta \\
&\qquad\quad+ \nu (\nu+1) \beta^2 \Big] \nabla^a \phi \nabla_a \phi \eqend{.}
\end{split}
\end{equations}

If we put an additional stress tensor on the right-hand side of~\eqref{eom_general}, which only has components in the extended directions $T_{mn}$, the reduced equations of motion~\eqref{eom} have the form
\begin{equations}[eom_stresstensor]
\begin{split}
&R_{mn} - \lambda (n+\nu-1)/\ell^2 \mathe^{2 \alpha \phi} g_{mn} - \alpha g_{mn} \nabla^a \nabla_a \phi \\
&\qquad- \left[ (n-2) \alpha + \nu \beta \right] \nabla_m \nabla_n \phi \\
&\qquad+ \left[ (n-2) \alpha^2 + 2 \nu \alpha \beta - \nu \beta^2 \right] \nabla_m \phi \nabla_n \phi \\
&\qquad- \alpha \left[ (n-2) \alpha + \nu \beta \right] g_{mn} \nabla^a \phi \nabla_a \phi \\
&\quad= 8 \pi G^{n+\nu}_\text{N} \left( T_{mn} - \frac{1}{(n+\nu-2)} g_{mn} T \right) \eqend{,}
\end{split} \\
\begin{split}
&\beta \Big[ \nabla^a \nabla_a \phi + \left[ (n-2) \alpha + \nu \beta \right] \nabla^a \phi \nabla_a \phi \Big] \\
&\qquad+ \left[ \lambda (n+\nu-1)/\ell^2 - k (\nu-1) H^2 \mathe^{- 2 \beta \phi} \right] \mathe^{2 \alpha \phi} \\
&\quad= \frac{8 \pi}{(n+\nu-2)} G^{n+\nu}_\text{N} T \eqend{,}
\end{split}
\end{equations}
where the trace is defined by $T = g^{mn} T_{mn}$.

\section{Curvature tensors for asymptotically AdS with a CHS}
\label{app_boundary}

For the metric~\eqref{aads_pert_metric} at $r=\text{const}$, we calculate to leading order in $h_{ab}$ and $\psi$
\begin{equations}
\Gamma^a_{bc} &= \frac{1}{2} \left( \partial_b h^a_c + \partial_c h^a_b - \partial^a h_{bc} \right) \eqend{,} \\
\Gamma^a_{b\gamma} &= 0 \eqend{,} \\
\Gamma^a_{\beta\gamma} &= \gamma^{(-1)}_{\beta\gamma} \eta \left( \delta^a_0 - \eta \partial^a \psi + 2 \delta^a_0 \psi + h^{a0} \right) \eqend{,} \\
\Gamma^\alpha_{bc} &= 0 \eqend{,} \\
\Gamma^\alpha_{b\gamma} &= \left( \eta^{-1} \delta^0_b + \partial_b \psi \right) \delta^\alpha_\gamma \eqend{,} \\
\Gamma^\alpha_{\beta\gamma} &= \Gamma^\alpha_{\beta\gamma}\left[\gamma^{(-1)}\right] \eqend{.}
\end{equations}
From this we obtain the Ricci tensor and scalar (using that the compact space is Einstein)
\begin{equations}
\begin{split}
\mathcal{R}_{bd} &= \partial^a \partial_{(b} h_{d)a} - \frac{1}{2} \partial^2 h_{bd} - \frac{1}{2} \partial_b \partial_d ( h + 2 \nu \psi ) \\
&\quad- \frac{\nu}{\eta} \left[ \partial_{(b} h_{d)0} - \frac{1}{2} \partial_\eta h_{bd} + 2 \delta^0_{(b} \partial_{d)} \psi \right] \eqend{,}
\end{split} \\
\mathcal{R}_{b\delta} &= 0 \eqend{,} \\
\begin{split}
\mathcal{R}_{\beta\delta} &= \gamma^{(-1)}_{\beta\delta} \bigg[ - \eta^2 \partial^2 \psi - \eta \partial^m h_{m0} + \frac{1}{2} \eta \partial_\eta ( h + 4 \nu \psi ) \\
&\qquad\quad+ (\nu-1) \left( 2 \psi + h_{00} \right) \bigg] \eqend{,}
\end{split} \\
\begin{split}
\mathcal{R} &= \frac{r^2}{\ell^2} \bigg[ \partial^m \partial^n h_{mn} - \partial^2 ( h + 2 \nu \psi ) - \frac{2 \nu}{\eta} \partial^m h_{m0} \\
&\qquad\quad+ \frac{\nu}{\eta} \partial_\eta ( h + 4 \nu \psi ) + \frac{\nu (\nu-1)}{\eta^2} \left( 2 \psi + h_{00} \right) \bigg] \eqend{.}
\end{split}
\end{equations}



\bibliographystyle{apsrev4-1}
\bibliography{references}

\end{document}